\documentclass[10pt,prd,preprint,floatfix,nofootinbib,superscriptaddress,showpacs,amssymb]{revtex4}
 \usepackage{graphicx}
\def\Z0{${\em Z^0\/}$}

\def\r#1 {$^{#1}$}

\newcommand{\gevc} { {\rm GeV/}c}
\newcommand{\gevcc}{ {\rm GeV/}c^2}

\def\gepsfcentered#1{
  \def\testit{#1}
  \def\lbracket{[}
  \ifx\testit\lbracket
    \let\dofilecmd=\gepsfwithopt
  \else
    \let\dofilecmd=\gepsfnoopt
  \fi
  \dofilecmd}

\def\gepsfnoopt#1{
  \begin{center}
  \leavevmode
  \epsffile{#1}
  \end{center}}

\def\gepsfwithopt#1 #2 #3 #4]#5{
  \begin{center}
  \leavevmode
  \gepsfmaxx=0.94\textwidth
  \epsffile[#1 #2 #3 #4]{#5}
  \end{center}}

\newdimen\gepsfmaxx
\def\epsfsize#1#2{
  \ifnum \epsfxsize=0
    \ifnum \epsfysize=0
      \ifnum #1 > \gepsfmaxx
        \gepsfmaxx
      \else
        #1
      \fi
    \else
      \epsfxsize
    \fi
  \else
    \epsfxsize
  \fi
}

\begin{document}
 \bibliographystyle{apsrev}
 \title{Search for narrow resonances lighter than $\Upsilon$ mesons}
\affiliation{Institute of Physics, Academia Sinica, Taipei, Taiwan 11529, Republic of China} 
\affiliation{Argonne National Laboratory, Argonne, Illinois 60439} 
\affiliation{University of Athens, 157 71 Athens, Greece} 
\affiliation{Institut de Fisica d'Altes Energies, Universitat Autonoma de Barcelona, E-08193, Bellaterra (Barcelona), Spain} 
\affiliation{Baylor University, Waco, Texas  76798} 
\affiliation{Istituto Nazionale di Fisica Nucleare Bologna, $^y$University of Bologna, I-40127 Bologna, Italy} 
\affiliation{Brandeis University, Waltham, Massachusetts 02254} 
\affiliation{University of California, Davis, Davis, California  95616} 
\affiliation{University of California, Los Angeles, Los Angeles, California  90024} 
\affiliation{University of California, San Diego, La Jolla, California  92093} 
\affiliation{University of California, Santa Barbara, Santa Barbara, California 93106} 
\affiliation{Instituto de Fisica de Cantabria, CSIC-University of Cantabria, 39005 Santander, Spain} 
\affiliation{Carnegie Mellon University, Pittsburgh, PA  15213} 
\affiliation{Enrico Fermi Institute, University of Chicago, Chicago, Illinois 60637}
\affiliation{Comenius University, 842 48 Bratislava, Slovakia; Institute of Experimental Physics, 040 01 Kosice, Slovakia} 
\affiliation{Joint Institute for Nuclear Research, RU-141980 Dubna, Russia} 
\affiliation{Duke University, Durham, North Carolina  27708} 
\affiliation{Fermi National Accelerator Laboratory, Batavia, Illinois 60510} 
\affiliation{University of Florida, Gainesville, Florida  32611} 
\affiliation{Laboratori Nazionali di Frascati, Istituto Nazionale di Fisica Nucleare, I-00044 Frascati, Italy} 
\affiliation{University of Geneva, CH-1211 Geneva 4, Switzerland} 
\affiliation{Glasgow University, Glasgow G12 8QQ, United Kingdom} 
\affiliation{Harvard University, Cambridge, Massachusetts 02138} 
\affiliation{Division of High Energy Physics, Department of Physics, University of Helsinki and Helsinki Institute of Physics, FIN-00014, Helsinki, Finland} 
\affiliation{University of Illinois, Urbana, Illinois 61801} 
\affiliation{The Johns Hopkins University, Baltimore, Maryland 21218} 
\affiliation{Institut f\"{u}r Experimentelle Kernphysik, Universit\"{a}t Karlsruhe, 76128 Karlsruhe, Germany} 
\affiliation{Center for High Energy Physics: Kyungpook National University, Daegu 702-701, Korea; Seoul National University, Seoul 151-742, Korea; Sungkyunkwan University, Suwon 440-746, Korea; Korea Institute of Science and Technology Information, Daejeon, 305-806, Korea; Chonnam National University, Gwangju, 500-757, Korea} 
\affiliation{Ernest Orlando Lawrence Berkeley National Laboratory, Berkeley, California 94720} 
\affiliation{University of Liverpool, Liverpool L69 7ZE, United Kingdom} 
\affiliation{University College London, London WC1E 6BT, United Kingdom} 
\affiliation{Centro de Investigaciones Energeticas Medioambientales y Tecnologicas, E-28040 Madrid, Spain} 
\affiliation{Massachusetts Institute of Technology, Cambridge, Massachusetts  02139} 
\affiliation{Institute of Particle Physics: McGill University, Montr\'{e}al, Qu\'{e}bec, Canada H3A~2T8; Simon Fraser University, Burnaby, British Columbia, Canada V5A~1S6; University of Toronto, Toronto, Ontario, Canada M5S~1A7; and TRIUMF, Vancouver, British Columbia, Canada V6T~2A3} 
\affiliation{University of Michigan, Ann Arbor, Michigan 48109} 
\affiliation{Michigan State University, East Lansing, Michigan  48824}
\affiliation{Institution for Theoretical and Experimental Physics, ITEP, Moscow 117259, Russia} 
\affiliation{University of New Mexico, Albuquerque, New Mexico 87131} 
\affiliation{Northwestern University, Evanston, Illinois  60208} 
\affiliation{The Ohio State University, Columbus, Ohio  43210} 
\affiliation{Okayama University, Okayama 700-8530, Japan} 
\affiliation{Osaka City University, Osaka 588, Japan} 
\affiliation{University of Oxford, Oxford OX1 3RH, United Kingdom} 
\affiliation{Istituto Nazionale di Fisica Nucleare, Sezione di Padova-Trento, $^z$University of Padova, I-35131 Padova, Italy} 
\affiliation{LPNHE, Universite Pierre et Marie Curie/IN2P3-CNRS, UMR7585, Paris, F-75252 France} 
\affiliation{University of Pennsylvania, Philadelphia, Pennsylvania 19104}
\affiliation{Istituto Nazionale di Fisica Nucleare Pisa, $^{aa}$University of Pisa, $^{bb}$University of Siena and $^{cc}$Scuola Normale Superiore, I-56127 Pisa, Italy} 
\affiliation{University of Pittsburgh, Pittsburgh, Pennsylvania 15260} 
\affiliation{Purdue University, West Lafayette, Indiana 47907} 
\affiliation{University of Rochester, Rochester, New York 14627} 
\affiliation{The Rockefeller University, New York, New York 10021} 
\affiliation{Istituto Nazionale di Fisica Nucleare, Sezione di Roma 1, $^{dd}$Sapienza Universit\`{a} di Roma, I-00185 Roma, Italy} 

\affiliation{Rutgers University, Piscataway, New Jersey 08855} 
\affiliation{Texas A\&M University, College Station, Texas 77843} 
\affiliation{Istituto Nazionale di Fisica Nucleare Trieste/Udine, I-34100 Trieste, $^{ee}$University of Trieste/Udine, I-33100 Udine, Italy} 
\affiliation{University of Tsukuba, Tsukuba, Ibaraki 305, Japan} 
\affiliation{Tufts University, Medford, Massachusetts 02155} 
\affiliation{Waseda University, Tokyo 169, Japan} 
\affiliation{Wayne State University, Detroit, Michigan  48201} 
\affiliation{University of Wisconsin, Madison, Wisconsin 53706} 
\affiliation{Yale University, New Haven, Connecticut 06520} 
\author{T.~Aaltonen}
\affiliation{Division of High Energy Physics, Department of Physics, University of Helsinki and Helsinki Institute of Physics, FIN-00014, Helsinki, Finland}
\author{J.~Adelman}
\affiliation{Enrico Fermi Institute, University of Chicago, Chicago, Illinois 60637}
\author{T.~Akimoto}
\affiliation{University of Tsukuba, Tsukuba, Ibaraki 305, Japan}
\author{B.~\'{A}lvarez~Gonz\'{a}lez$^t$}
\affiliation{Instituto de Fisica de Cantabria, CSIC-University of Cantabria, 39005 Santander, Spain}
\author{S.~Amerio$^z$}
\affiliation{Istituto Nazionale di Fisica Nucleare, Sezione di Padova-Trento, $^z$University of Padova, I-35131 Padova, Italy} 

\author{D.~Amidei}
\affiliation{University of Michigan, Ann Arbor, Michigan 48109}
\author{A.~Anastassov}
\affiliation{Northwestern University, Evanston, Illinois  60208}
\author{A.~Annovi}
\affiliation{Laboratori Nazionali di Frascati, Istituto Nazionale di Fisica Nucleare, I-00044 Frascati, Italy}
\author{J.~Antos}
\affiliation{Comenius University, 842 48 Bratislava, Slovakia; Institute of Experimental Physics, 040 01 Kosice, Slovakia}
\author{G.~Apollinari}
\affiliation{Fermi National Accelerator Laboratory, Batavia, Illinois 60510}
% ...

\collaboration{CDF Collaboration\footnote{With visitors from $^a$University of Massachusetts Amherst, Amherst, Massachusetts 01003,
$^b$Universiteit Antwerpen, B-2610 Antwerp, Belgium, 
$^c$University of Bristol, Bristol BS8 1TL, United Kingdom,
$^d$Chinese Academy of Sciences, Beijing 100864, China, 
$^e$Istituto Nazionale di Fisica Nucleare, Sezione di Cagliari, 09042 Monserrato (Cagliari), Italy,
$^f$University of California Irvine, Irvine, CA  92697, 
$^g$University of California Santa Cruz, Santa Cruz, CA  95064, 
$^h$Cornell University, Ithaca, NY  14853, 
$^i$University of Cyprus, Nicosia CY-1678, Cyprus, 
$^j$University College Dublin, Dublin 4, Ireland,
$^k$University of Edinburgh, Edinburgh EH9 3JZ, United Kingdom, 
$^l$University of Fukui, Fukui City, Fukui Prefecture, Japan 910-0017
$^m$Kinki University, Higashi-Osaka City, Japan 577-8502
$^n$Universidad Iberoamericana, Mexico D.F., Mexico,
$^o$University of Iowa, Iowa City, IA  52242,
$^p$Queen Mary, University of London, London, E1 4NS, England,
$^q$University of Manchester, Manchester M13 9PL, England, 
$^r$Nagasaki Institute of Applied Science, Nagasaki, Japan, 
$^s$University of Notre Dame, Notre Dame, IN 46556,
$^t$University de Oviedo, E-33007 Oviedo, Spain, 
$^u$Texas Tech University, Lubbock, TX  79609, 
$^v$IFIC(CSIC-Universitat de Valencia), 46071 Valencia, Spain,
$^w$University of Virginia, Charlottesville, VA  22904,
$^x$Bergische Universit\"at Wuppertal, 42097 Wuppertal, Germany,
$^{ff}$On leave from J.~Stefan Institute, Ljubljana, Slovenia, 
}}
\noaffiliation

\noaffiliation
 \begin{abstract}
   We report a search for narrow resonances, produced in $p\bar{p}$ collisions
   at $\sqrt{s}=1.96$ TeV, that decay into muon pairs with invariant mass
   between 6.3 and 9.0 $\gevcc$. The data, collected with the CDF~II detector
   at the Fermilab Tevatron collider, correspond to an integrated luminosity
   of 630 pb$^{-1}$. We use the dimuon invariant mass distribution to set
   90\% upper credible limits of about 1\% to the ratio of the production 
   cross section times muonic branching fraction of possible narrow resonances
   to that of the $\Upsilon(1{\rm S})$ meson.\\
 \end{abstract}
 \preprint{FERMILAB-PUB-09-067-E}
 \maketitle
 \section {Introduction} \label{sec:ss-intro}
   Searches for narrow resonances at SPEAR have set upper bounds on $\Gamma_l$,
   the leptonic width of possible resonances, of $100$ eV in the cms energy
   region $5.7\leq E_{\rm cm}\leq 6.4$ GeV and of approximately $60$ eV in the
   region $7.0\leq E_{\rm cm}\leq 7.4$ GeV~\cite{spear}. These limits do not
   exclude bound states of charge $-1/3$ and spin $0$ light quarks,  the
   existence of which is required by supersymmetric theories. The leptonic
   width of {\Large $\varepsilon$} mesons, 1$^{--}$ bound states of charge
   $-1/3$ scalar quarks, has been evaluated in Ref.~\cite{nappi} using
   potential models of ordinary heavy quarks~\cite{pot-mod}. Because of the 
   p-wave suppression of the fermion contribution, the leptonic width is
   estimated to be approximately $18$ ($6$) eV for a resonance with a $6$ 
   ($10$) $\gevcc$ mass. 
   A more recent study~\cite{dimu_1} has set an average 90\% upper credible
   limit of $\Gamma_l \simeq 8$ eV to the existence of narrow resonances in the
   invariant  mass range $6.3-9 \, \gevcc$ by using a dimuon sample 
   corresponding to a luminosity of approximately 110 pb$^{-1}$ collected with
   the CDF detector during the $1992-1995$ Tevatron collider run. Because of 
   statistical fluctuations, that study does not rule out the existence of 
   {\Large $\varepsilon$} states over the full mass range.
   In particular, at a mass of $7.2 \, \gevcc$, that study returns an upper
   credible limit that is consistent with what is expected if an 
   {\Large $\varepsilon$} state  is produced.
    
   In this study, we repeat the analysis in  Ref.~\cite{dimu_1} and take 
   advantage of a higher statistics data sample that corresponds to an
   integrated luminosity of 630 pb$^{-1}$ collected with the CDF II detector
   after May 2006. At that time, the CDF II trigger system was upgraded and
   made capable of acquiring events containing muon pairs with invariant mass
   larger than $6 \; \gevcc$ and a kinematic acceptance comparable to that of
   the Run I trigger system. Section~\ref{det} describes the detector systems
   relevant to this analysis. The data sample and analysis method are described
   in Sec.~\ref{sec:ss-mudat}. In Sec.~\ref{sec:ss-mueps}, we fit the dimuon
   invariant mass distribution and derive $90$\%  upper credible limits to
   $\Gamma_l$ as a function of the resonance mass.
   Our conclusions are summarized in Sec.~\ref{sec:concl}.
 \section{CDF~II detector and trigger system} \label{det}
  CDF II is a multipurpose detector, equipped with a charged particle
  spectrometer and a finely segmented calorimeter. In this section, we briefly
  describe the detector components that are relevant to this analysis. A
  detailed description of these subsystems can be found in
  Refs.~\cite{det1,det2,det3_0,det3,det4_0,det4,det5,det6,det7,det8}. Two
  devices inside the 1.4 T solenoid are used for measuring the momentum of
  charged particles: the silicon vertex detector (SVXII and ISL) and the
  central tracking chamber (COT). The SVXII detector consists of microstrip 
  sensors arranged in six cylindrical shells with radii between 1.5 and
  10.6 cm, and with a total $z$ coverage~\cite{coord} of 90 cm. The first
  SVXII layer, also referred to as the L00 detector, is made of single-sided 
  sensors mounted on the beryllium beam pipe. The remaining five SVXII layers
  are made of double-sided sensors and are divided into three contiguous
  five-layer sections along the beam direction $z$. The two additional silicon
  layers of the ISL help to link tracks in the COT to hits in the SVXII. 
  The COT is a cylindrical drift chamber containing 96 sense wire layers
  grouped into eight alternating superlayers of axial and stereo wires. Its
  active volume covers $|z| \leq 155$ cm and 40 to 140 cm in radius. The
  transverse momentum resolution of tracks reconstructed using COT hits
  is $\sigma(p_T)/p_T^2 \simeq 0.0017\; [\gevc]^{-1}$.

  The central muon detector (CMU) is located around the central electromagnetic
  and hadronic calorimeters, which have a thickness of 5.5 interaction lengths
  at normal incidence. The CMU detector covers a nominal pseudorapidity range
  $|\eta| \leq 0.63$ relative to the center of the detector, and  is segmented
  into two barrels of 24 modules, each covering 15$^\circ$ in $\phi$. Every
  module is further segmented  into three  submodules, each covering
  4.2$^\circ$ in $\phi$ and consisting of four layers of drift chambers. The
  smallest drift unit, called a stack, covers a 1.2$^\circ$ angle in $\phi$.
  Adjacent pairs of stacks are combined together into a tower. A track segment
  (hits in two out of four layers of a stack) detected in a tower is referred
  to as a CMU stub. A second set of muon drift chambers (CMP) is located behind
  an additional steel absorber of 3.3 interaction lengths. The chambers are
  640 cm long and are arranged axially to form a box around the central 
  detector. The CMP detector covers a nominal pseudorapidity range 
  $|\eta| \leq 0.54$ relative to the center of the detector. Muons that
  produce a stub in both CMU and CMP systems are called CMUP muons.
  The CMX muon detector consists of eight drift chamber layers and
  scintillation counters positioned behind the hadron calorimeter. The CMX
  detector extends the muon coverage to $|\eta| \leq 1$.

  The luminosity is measured using gaseous Cherenkov counters (CLC) that
  monitor the rate of inelastic $p\bar{p}$ collisions. The inelastic $p\bar{p}$
  cross section at $\sqrt{s}=1960$ GeV is scaled from measurements at 
  $\sqrt{s}=1800$ GeV using the calculations in Ref.~\cite{sigmatot}. The
  integrated luminosity is determined with a 6\% systematic
  uncertainty~\cite{klimen}.

  CDF uses a three-level trigger system. At Level 1 (L1), data from every beam 
  crossing are stored in a pipeline capable of buffering data from 42 beam 
  crossings. The L1 trigger either rejects events or copies them into one of 
  the four Level 2 (L2) buffers. Events that pass the L1 and L2 selection 
  criteria are sent to the Level 3 (L3) trigger, a cluster of computers 
  running  speed-optimized reconstruction code.  
  For this study, we select events with two muon candidates identified by the
  L1 and L2 triggers. The L1 trigger uses tracks with $p_T \geq 1.5 \; \gevc$
  found by a fast track processor (XFT). The XFT examines COT hits from the
  four axial superlayers and uses the $r-\phi$ information to find tracks with
  $p_T \geq 1.5 \; \gevc$ in azimuthal sections of 1.25$^\circ$. The XFT passes
  the tracks to a set of extrapolation units that determine the CMU (CMX)
  towers in which a CMU (CMX) stub  should be found if the track is a muon.
  If a stub is found, a L1 CMU (CMX) primitive is generated.
  The L1 dimuon trigger requires one CMU primitive and an additional CMU or CMX
  primitive. The L2 trigger additionally requires that at least one of the
  muons has a CMUP stub matched to an XFT track with $p_T \geq 3 \;\gevc$ and
  that  additional CMU (CMX) stubs are matched to  XFT tracks with 
  $p_T \geq 1.5\, (2) \;\gevc$. The L2 trigger is dynamically prescaled as a
  function of the Tevatron luminosity in order to optimize the data acquisition
  performance. The L3 trigger selects  muon pairs with  invariant mass larger
  than $5 \; \gevcc$, and $|\delta z_0| \leq 5$ cm, where $z_0$ is the $z$ 
  coordinate of the muon track at its point of closest approach to the beam 
  line in the $r-\phi$ plane. These requirements define the dimuon trigger
  used in this analysis.
 \section{Dimuon data sample} \label{sec:ss-mudat} 
  In this analysis, we select events acquired with the dimuon trigger and
  which contain two and only two muons with opposite charge. Events are
  reconstructed offline taking advantage of more refined calibration constants
  and reconstruction algorithms. COT tracks are extrapolated into the SVXII
  detector, and refitted adding silicon hits  consistent  with the track
  extrapolation into the microvertex detector. We require that at least one
  muon has stubs reconstructed in both the CMU and CMP detectors and is matched
  to a track with $p_T \geq 3 \; \gevc$. A track is identified as a CMUP muon
  if $\Delta r\phi$, the distance in the $r-\phi$ plane between the track
  projected to the CMU (CMP) chambers and a CMU (CMP) stub, is less than 30 
  (40) cm. The additional muon is accepted if  a stub reconstructed in the CMU
  or CMX detectors is matched to a track with $p_T \geq 2 \; \gevc$. A track is
  identified as a CMU (CMX) muon if $\Delta r\phi$ is less than 30 (50) cm.
  We require that muon-candidate stubs correspond to  L1 primitives, and
  correct the muon momentum for energy losses in the detector.
 
  Previous measurements of the $B$ hadron mixing probability~\cite{bmix} and 
  of the $b\bar{b}$ correlated production~\cite{bbarbxs} have used similarly
  selected dimuon data samples. Those studies show that $70-75$\% of these
  dimuon pairs are contributed by $b\bar{b}$ and $c\bar{c}$ production.
  Following the procedure pioneered in Ref.~\cite{dimu_1}, we reject muons 
  arising from the decay of hadrons with heavy flavor with the following
  requirements:
 \begin{enumerate}
 \item Isolation. The isolation, $I$, is defined as the scalar
       sum of the transverse momenta of all the tracks in a cone of radius
       $R = \sqrt{\delta\phi^2 + \delta\eta^2}=0.4$ around the muon direction.
       We require that both muons have isolation $I \leq 4\; \gevc$.
 \item Promptness. In contrast to $b$ and $c$ hadrons, the
       $\Upsilon(1{\rm S})$ and {\Large $\varepsilon$} mesons have a negligible
       lifetime. We evaluate the invariant mass of each muon pair  
       by constraining the two muon tracks to originate from a common point
       in the three-dimensional space (vertex constraint). We reject
       muon pairs if the probability of the vertex-constrained fit is smaller
       than 0.001. We also reject muon pairs if $L_{xy}/\sigma_{L_{xy}}\geq 3$,
       where $L_{xy}$ is the displacement of the dimuon-candidate vertex with
       respect to the $p\bar{p}$ event vertex projected onto the dimuon 
       transverse momentum vector, and $\sigma_{L_{xy}}$ is its uncertainty.
 \end{enumerate}
  The invariant mass distribution of the muon pairs passing our selection
  requirements is shown in Fig.~\ref{fig:fig_dimu_1}.
\section{Search for narrow resonances} \label{sec:ss-mueps}
  As in Ref.~\cite{dimu_1}, we limit our search for narrow resonances to the
  mass region above $6.3\; \gevcc$, where the kinematic acceptance becomes
  independent of the {\Large $\varepsilon$}-candidate transverse momentum. 
 \begin{figure}
 \begin{center}
 \vspace{-0.2in}
 \leavevmode
 \includegraphics*[width=0.5\textwidth]{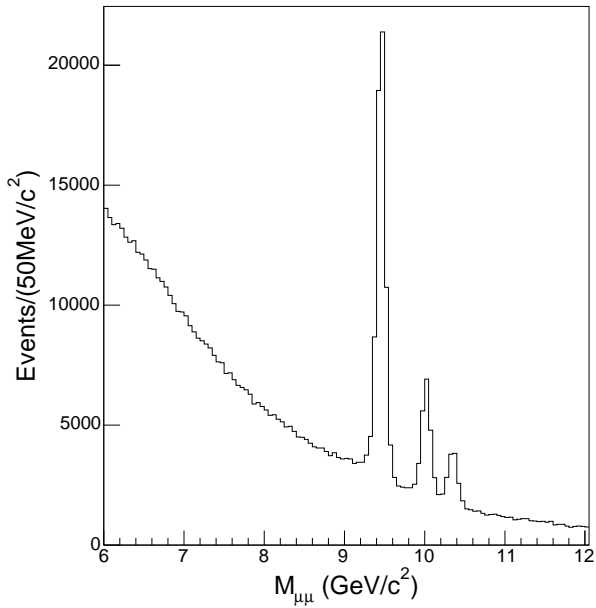}
 \caption[]{Invariant mass distribution of the selected muon pairs. }
 \label{fig:fig_dimu_1}
 \end{center}
 \end{figure}
  The number of $\Upsilon$ mesons in the data is derived by fitting the
  invariant mass distribution in Fig.~\ref{fig:fig_dimu_1} with a binned 
  maximum likelihood method~\cite{minuit}. We use a fifth-order polynomial
  to model the continuum in the invariant mass region $6-12 \; \gevcc$ and
  Gaussian functions to model the $\Upsilon$ contributions~\footnote{
  Two Gaussian functions are required to model each $\Upsilon(1{\rm S})$ or
  $\Upsilon(2{\rm S})$ signal, whereas a single Gaussian function is sufficient
  to describe the $\Upsilon(3{\rm S})$ signal.}.  
  The best fit returns $52700\pm 350\; \Upsilon(1{\rm S})$ meson events.
  The fit also returns $M_{\Upsilon({\rm 1S})} = 9459 \pm 1$ MeV/$c^2$ and
  a mass resolution $\sigma_{M} = 52 \pm 1$ MeV/$c^2$. This mass resolution is
  well modeled by a simulation of the process 
  $p\bar{p}\rightarrow \Upsilon({\rm 1S}) X $. In the simulation, the event 
  generator produces unpolarized $\Upsilon({\rm 1S})$ mesons with the
  transverse momentum distribution of the data~\cite{upsi-sec} and a flat
  rapidity distribution for $|y| \leq 1$. The generated events are processed 
  with the CDF~II detector simulation that in turn is based on the {\sc geant}
  simulation package~\cite{geant}. Events are then required to pass the same
  trigger requirements, analysis selection, and reconstruction criteria imposed
  on the data. The same simulation~\footnote{
  When simulating {\Large $\varepsilon$} states, the transverse momentum
  distribution is rescaled from that of the $\Upsilon({\rm 1S})$ data so that
  $<p_T^{\mbox{\Large $\varepsilon$}}>/<p_T^{\Upsilon({\rm 1S})}>=
  m_{\mbox{\Large $\varepsilon$}}/m_{\Upsilon({\rm 1S})}$. 
  For $J/\psi$ mesons, this rescaling procedure predicts a $d\sigma/dp_T$
  distribution that decreases more rapidly with increasing momenta than the
  distribution of the data. However, a poor modeling of the transverse
  momentum distribution is not a cause of significant error because the
  kinematic acceptance does not depend on the  {\mbox{\Large $\varepsilon$}}
  transverse momentum within a couple of percent.}
  predicts an invariant mass resolution that increases from 32  MeV/$c^2$ for
  an {\Large $\varepsilon$} state with a mass of 6.3 $\gevcc$ to 50 MeV/$c^2$ 
  for $m_{\mbox{\Large $\varepsilon$}}= 9 \; \gevcc$.
 
  The invariant mass distribution in Fig.~\ref{fig:fig_dimu_1} does not exhibit
  any significant narrow bump in the mass range $6.3-9 \;\gevcc$.
  We set limits to
   $R= [\sigma_{\mbox{\Large $\varepsilon$}}
     \; BR(\mbox{{\Large $\varepsilon$}} \rightarrow \mu\mu)]/ 
       [\sigma_{\Upsilon({\rm 1S})}
      \;  BR(\Upsilon({\rm 1S}) \rightarrow \mu \mu)]$,
  the production cross section times the muonic branching fraction, under the
  assumptions that {\mbox{\Large $\varepsilon$}} states have negligible widths
  with respect to the experimental resolution and are produced unpolarized.
  We also assume that the efficiency of the isolation requirement, used to
  suppress the heavy flavor contribution, does not depend on the
  {\mbox{\Large $\varepsilon$}} mass and is equal to that for
  $\Upsilon(1{\rm S})$ mesons. Limits on the existence of narrow
  {\Large $\varepsilon$} states are set by adding a Gaussian term to the
  likelihood function used to fit the data in the $6-12 \; \gevcc$ mass
  interval. We perform $108$ fits in which we change the position of the
  Gaussian peak in steps of 25  MeV/$c^2$ from 6.3 to 9.0 $\gevcc$.
  In each fit, we force the Gaussian width to the rms mass resolution predicted
  by the detector simulation for that mass. The usage of only one Gaussian
  function to estimate the number of {\mbox{\Large $\varepsilon$}} events 
  introduces a 4.5\% systematic uncertainty. The uncertainty is estimated as
  the difference between the number of $\Upsilon(1{\rm S})$ events returned
  by fits that use one or two Gaussian functions to estimate the 
  $\Upsilon(1{\rm S})$ signal. For each {\mbox{\Large $\varepsilon$}} mass
  value, we use the integral of the Gaussian function and its error returned 
  by the best fit to derive $N_{ul}$, the $90$\% credibility upper limit to 
  the number of events contributed by a narrow resonance~\footnote{ 
   The integral of the fit likelihood from $N_{ul}$ to infinity is $10$\% of 
   the integral of the fit likelihood from $0$ to infinity. The justification
   for this procedure is Bayesian with a prior that is zero for negative 
   resonance cross sections and flat for positive ones.}.
  We evaluate the ratio of the geometric and kinematic acceptance for an 
  {\mbox{\Large $\varepsilon$}} resonance to that for the $\Upsilon(1{\rm S})$
  meson with the previously described Monte Carlo simulation. The geometric and
  kinematic acceptance depends on the resonance mass and increases from 65.5\%
  of that of the  $\Upsilon(1{\rm S})$ meson for
  $m_{\mbox{\Large $\varepsilon$}}=6.3 \; \gevcc$ to 97.4\% 
  for  $m_{\mbox{\Large $\varepsilon$}}=9.0 \; \gevcc$.
  Small data-to-simulation correction factors for the trigger and
  reconstruction efficiency~\cite{bbarbxs,bjk} depend little on the muon $p_T$
  and are neglected  as in the study in Ref.~\cite{dimu_1}. The kinematic
  acceptance  for $\Upsilon(1{\rm S})$ mesons depends on their polarization.
  If $\Upsilon(1{\rm S})$ mesons are produced longitudinally polarized, the
  acceptance increases by 5\%, whereas it decreases by 3\% if
  $\Upsilon(1{\rm S})$ mesons are produced  transversely polarized.
  The current knowledge of the  polarization of $\Upsilon(1{\rm S})$ mesons
  produced at the Tevatron~\cite{Acosta:2001gv,UpsiPolarizationD0} yields 
  an additional 3\% systematic uncertainty on the ratio of the
  {\mbox{\Large $\varepsilon$}} to $\Upsilon(1{\rm S})$ production cross 
  sections. The total systematic uncertainty on the ratio (6\%) is added in
  quadrature to the statistical errors returned by the fits. 

\begin{figure}
 \begin{center}
 \vspace{-0.2in}
 \leavevmode
 \includegraphics*[width=0.5\textwidth]{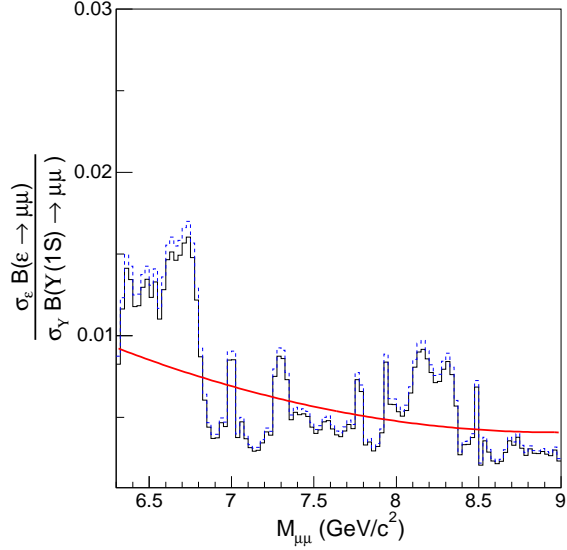}
 \caption[]{Bayesian $90$\% upper credible limits (histogram) to 
            $\frac { \sigma_{\mbox{\Large $\varepsilon$}}
            \; BR(\mbox{{\Large $\varepsilon$}} \rightarrow \mu\mu)}
            {\sigma_{\Upsilon({\rm 1S})}\;  
            BR(\Upsilon({\rm 1S}) \rightarrow \mu \mu)}$
            as a function of the dimuon invariant mass.
            The dashed histogram represents limits that include the 6\%
            systematic uncertainty. The solid line represents the limits
            expected when  fits return no {\mbox{\Large $\varepsilon$}}
            events.}
 \label{fig:fig_lim1}
 \end{center}
 \end{figure}

\begin{figure}
 \begin{center}
 \vspace{-0.2in}
 \leavevmode
 \includegraphics*[width=0.445\textwidth]{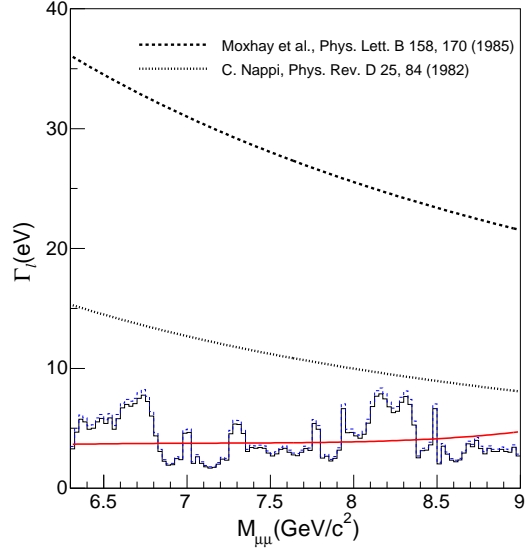}
 \caption[]{Bayesian $90$\% upper limits to 
            $\Gamma_l^{\mbox{\Large $\varepsilon$}}$ (histogram).
            The dashed histogram represents  limits that include the 6\% 
            systematic uncertainty. The solid line represents the limits
            expected when fits return no {\mbox{\Large $\varepsilon$}} events.
            The dashed and dotted lines represent the leptonic widths of
            1$^{--}$ bound states of scalar quarks predicted in 
            Refs.~\cite{nappi} and~\cite{moxhay}, respectively.}
 \label{fig:fig_lim2}
 \end{center}
 \end{figure}

  The $90$\% credibility upper limit to $R$ is given by the ratio of $N_{ul}$
  to the number of observed $\Upsilon(1{\rm S})$ mesons, corrected for the 
  relative acceptance. These limits are shown in Fig.~\ref{fig:fig_lim1} as a
  function of the ${\mbox{\Large $\varepsilon$}}$ mass.
 Figure~\ref{fig:fig_lim2} shows the $90$\%  upper credible limits to
  $\Gamma_l^{\mbox{\Large $\varepsilon$}}$. The latter limits are derived
  under the assumption~\footnote{
    As noted in Ref.~\cite{dimu_1}, this equation tends to underestimate the
    production cross section of states lighter than the $\Upsilon(1{\rm S})$
    meson. If  this were the case, the $\Gamma_l^{\mbox{\Large $\varepsilon$}}$
    limits set by our study would be correspondingly smaller than those
    indicated in Fig.~\ref{fig:fig_lim2}.}
  that $\sigma_{\mbox{\Large $\varepsilon$}}\;
        BR(\mbox{\Large $\varepsilon$} \rightarrow \mu\mu) =
        \sigma_{\Upsilon({\rm 1S})} \; 
        BR(\Upsilon({\rm 1S}) \rightarrow \mu \mu)  \times (m_{\Upsilon(1S)}/
        m_{\mbox{\Large $\varepsilon$}})^3 
       \times \Gamma_{\mu }^{\mbox{\Large $\varepsilon$}}/ 
       \Gamma_{\mu }^{\Upsilon({\rm 1S})}$~\cite{dimu_1}.

 \section{Conclusions} \label{sec:concl}
   We present a study of the invariant mass spectrum of muon pairs produced at
   the Fermilab Tevatron collider and collected by the CDF experiment. This
   study improves previous limits to the existence of narrow resonances in the
   invariant mass range $6.3-9.0 \; \gevcc$. We derive  90\% upper credible
   limits to $R$, the ratio of the production cross section times muonic
   branching fraction of possible narrow resonances to that of the
   $\Upsilon(1{\rm S})$ meson, of the order of 1\%. Assuming that 
   $\sigma_{\mbox{\Large $\varepsilon$}} \; BR(\mbox{{\Large $\varepsilon$} })=
    \sigma_{\Upsilon({\rm 1S)}} \;  BR(\Upsilon({\rm 1S}) \rightarrow \mu \mu) 
        \times (m_{\Upsilon({\rm 1S})}/ m_{\mbox{\Large $\varepsilon$}})^3 
        \times \Gamma_{\mu }^{\mbox{\Large $\varepsilon$}}/ 
        \Gamma_{\mu }^{\Upsilon({\rm 1S})}$,
   these limits correspond to  $90$\% upper credible limits smaller than $10$
   eV to the leptonic width of possible narrow resonances.

 \section{Acknowledgments}
  We thank the Fermilab staff and the technical staffs of the participating 
  institutions for their vital contributions. This work was supported by the
  U.S. Department of Energy and National Science Foundation; the Italian 
  Istituto Nazionale di Fisica Nucleare; the Ministry of Education, Culture,
  Sports, Science and Technology of Japan; the Natural Sciences and Engineering
  Research Council of Canada; the National Science Council of the Republic of 
  China; the Swiss National Science Foundation; the A.P. Sloan Foundation; the
  Bundesministerium f\"ur Bildung und Forschung, Germany; the Korean Science 
  and Engineering Foundation and the Korean Research Foundation; the Science
  and Technology Facilities Council and the Royal Society, UK; the Institut
  National de Physique Nucleaire et Physique des Particules/CNRS; the Russian
  Foundation for Basic Research; the Ministerio de Ciencia e Innovaci\'{o}n,
  and Programa Consolider-Ingenio 2010, Spain; the Slovak R\&D Agency; and the
  Academy of Finland. 




\begin{thebibliography}{99}
\bibitem{coord}  In the CDF coordinate system, $\theta$ and $\phi$ are the polar and azimuthal
 angles of a track, respectively, defined with respect to the proton beam
 direction, $z$. The pseudorapidity $\eta$ is defined as 
 $-\ln \;\tan (\theta/2)$. The transverse momentum of a particle is 
 $p_T= p \; \sin (\theta)$. The rapidity is defined as 
 $y=1/2 \cdot \ln ( (E+p_z)/(E-p_z) )$, where $E$ and $p_z$ are the energy
 and longitudinal momentum of the particle associated with the track, respectively.
\end{thebibliography}
 \end{document}